\documentstyle[psfig,twocolumn]{esa_sp_latex}

\newcommand{\Inuk}{\mbox{$I_{\nu}(\vec{k})$}}
\newcommand{\vk}{\mbox{$\vec{k}$}}
\newcommand{\beq}{\begin{equation}}
\newcommand{\eeq}{\end{equation}}
\newcommand{\beqar}{\begin{eqnarray}}
\newcommand{\eeqar}{\end{eqnarray}}

\begin{document}
%

\parindent 0pt
\parskip 10pt plus 1pt minus 1pt
\hoffset=-1.5truecm
\topmargin=-1.0cm
\textwidth 17.1truecm \columnsep 1truecm \columnseprule 0pt 

\title{\bf AN ANISOTROPIC ILLUMINATION MODEL OF SEYFERT I GALAXIES}

\author{{\bf P.O.~Petrucci, G.~Henri} \vspace{2mm} \\
Laboratoire d'Astrophysique, Observatoire de Grenoble, B.P 53X, F38041
Grenoble Cedex, France}

\maketitle

\begin{abstract}

 We present a new model of accretion disk where the disk luminosity is 
 entirely due to the reprocessing of hard radiation impinging on the disk. 
 The hard radiation itself is emitted by a hot point source above the
 disk, that could be physically realized by a strong shock terminating an
 aborted  
 jet. This hot source contains ultra-relativistic leptons scattering the disk
 soft photons by Inverse Compton (IC) process. Using simple formula to 
 describe the IC process in an anisotropic photon field, we derive a 
 self-consistent solution in the Newtonian geometry, where the angular 
 distribution of soft and hard radiation, and 
 the radial profile of the disk effective temperature are determined in a
 univocal way. This offers an alternative picture to the standard accretion 
 disk emission law, reproducing individual spectra and 
 predicting new scaling laws that fit better the observed statistical 
 properties. General relativistic calculations are also carried
 out. It appears that differences with the Newtonian case are weak,
 unless the hot source is very close to the black hole.    \vspace {5pt} \\

  Keywords: black hole; galaxies: nuclei; galaxies: Seyfert; relativity

\end{abstract}

\section{INTRODUCTION}
It is widely believed that the high energy emission of AGNs is produced
by Comptonization of soft photons by high energy electrons or pairs.
Besides, for Seyfert galaxies, some observational facts support 
the idea that high energy radiation can be primarily produced and 
reflected on a cold surface, producing a fair fraction of thermal 
UV-optical radiation (\cite*{Cla92},\cite*{Pou90}). We then  propose a new
model involving a point source of relativistic leptons located 
above the disk (that could be physically realized by a
strong shock terminating an aborted jet)
emitting hard radiation by Inverse Compton (IC) process on soft photons
produced by the accretion disk. The disk itself radiates only through the
re-processing of the hard radiation impinging on it, i.e. we do not suppose
any internal energy dissipation (this is a relatively good approximation
since the disk supplies its jet with almost all the available gravitational
power, being then weakly dissipative (\cite*{Ferr})) . Such a geometry is
highly anisotropic, which takes a real importance in the computation of IC
process (\cite*{Ghi91}, \cite*{Hen*}). We treat both Newtonian and general
relativistic cases, deriving a self-consistent solution in the Newtonian
case. We present here
the main equations describing the radiative balance
between the hot source and the disk, as well as the
most important results supplied by the model.

\section{The model}

\begin{figure*}
  \begin{center}
  \psfig{width=17cm,height=4cm,angle=-90,file=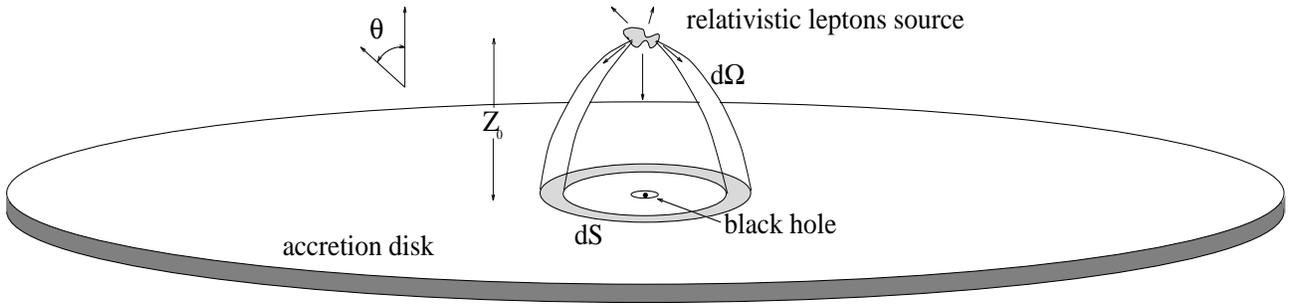}
  \caption{\em The general picture of the model. We have also drawn the
    trajectory of a beam of photons  emitted by the hot source in a solid
    angle $d\Omega$ and absorbed by a surface ring $dS$ on the
    disk.}
  \end{center}
  \label{main}
\end{figure*}
Let us consider a relativistic charged
particle, characterized by its the Lorentz factor $\gamma=(1-\beta 
^2)^{-1/2}$, and the soft photon field, characterized by the intensity
distribution $I_{\nu}(\vec{k} )$. We assume that the Thomson approximation
is valid; in this limit, the rate of energy emitted by the particle by
inverse Compton process is: 
\begin{equation}
dP \propto \gamma^2 \int \Inuk(1-\beta \vec{k_0} . \vec{k})^2 d\Omega d\nu
\end{equation}
$\vk$ and $\vk_0$ are respectively the unit vectors along the photon and the
particle velocity.
Thus, one deduced with the hypothesis of an isotropic distribution of
high energy particles, the plasma emissivity of the hot
source (\cite*{Hen*}):
\begin{equation}
\frac{dP}{d\Omega} \propto [(3J-K)-4H\mu +(3K-J)\mu^2] \label{eqdP}
\end{equation}
where the 3
Eddington parameters, J, H and K are defined by:
\beqar
J & = & {1 \over 2}\int \Inuk d\mu d\nu \\
H & = & {1 \over 2}\int \Inuk \mu d\mu d\nu\\
K & = & {1 \over 2}\int \Inuk \mu^2 d\mu d\nu
\eeqar
and  $\mu = \cos\theta$ is the cosine of the impinging angle of radiation
(cf Fig. \ref{main}). Under the hypothesis that the disk reprocesses the
whole radiation 
impinging on it, we equalize the power absorbed and emitted by a 
surface element $dS$ of the disk at a distance $r$ of the black hole (cf
Fig. \ref{main}): 
\begin{eqnarray}
      F(r) dS &=& \frac {dP}{d\Omega}d\Omega = -\mu^3 \frac {dP}{d\Omega}\frac 
      {dS}{Z_0^2} \label{eqF}\\
              &=&\pi I(r)dS \label{eqI}
\end{eqnarray}
where $Z_0$ is the height of the hot source above the disk (cf
Fig. \ref{main}). 
In equation (\ref{eqI}), one supposes that the disk radiates like a black
body. Finally, combining Equations (\ref{eqdP}), (\ref{eqF}) and
(\ref{eqI}), one obtain a linear system of equations between $J,\ H$ and
$K$. By setting 
its determinant to zero, we finally find universal solutions for the hot source
and disk emissivity laws. The relativistic case of a Kerr metrics has
also been solved (\cite{Pet*}). In this case, the overall spectra depend
on $a$, the 
angular momentum by unit mass of the black hole, and on the ratio $Z_0/M$
of the height of the hot source on the black hole mass. The whole
computations in the Newtonian and Kerr geometry can be found in
(\cite*{Hen*}) and (\cite*{Pet*}).   

\section{Results}
\subsection{Angular distribution of the hot source}
\begin{figure}
  \psfig{width=7.cm,angle=90,height=7.cm,file=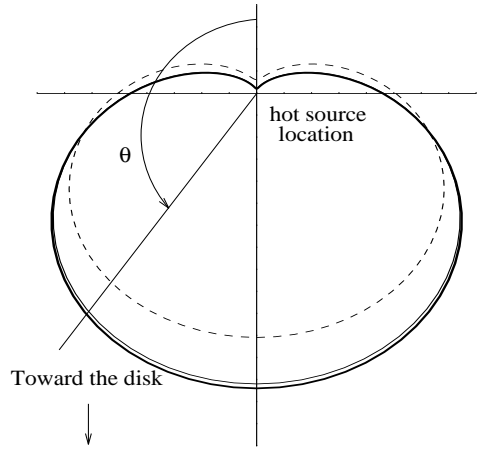}
  \caption[]{\em Polar plots of $\displaystyle\frac{dP}{d\Omega}$ for
     $Z_0/M=100$ (solid line) and $Z_0/M=10$ (dashed line) in Kerr
     metrics with $a=0.998$. The bold line corresponds to the Newtonian
     metrics}
  \label{anisotropy}
\end{figure}

It appears from Eq. (\ref{eqdP}) that the anisotropy of the soft photon
field about level with the hot source, leads to an anisotropic Inverse
Compton process, with much more radiation being scattered backward than
forward. Such a
anisotropic re-illumination could naturally explain the apparent X-ray
luminosity, usually much lower than the optical-UV continuum emitted in
the blue bump. It can also explain the equivalent width observed for the
iron line, which requires more impinging radiation than what is actually
observed. 
 We plot in
Figure \ref{anisotropy} the angular distribution of the power emitted by
the hot source in Newtonian metrics and for different values of the
source height in Kerr metrics. It appears that
the closer the source to the black hole is, the less anisotropic the
photon field is. This is principally due to the curvature of geodesics
making the photons emitted near the black hole arrive at larger angle
than in the Newtonian case. 

\subsection{Disk temperature profile}
\begin{figure}
 \psfig{width=8.cm,height=6cm,file=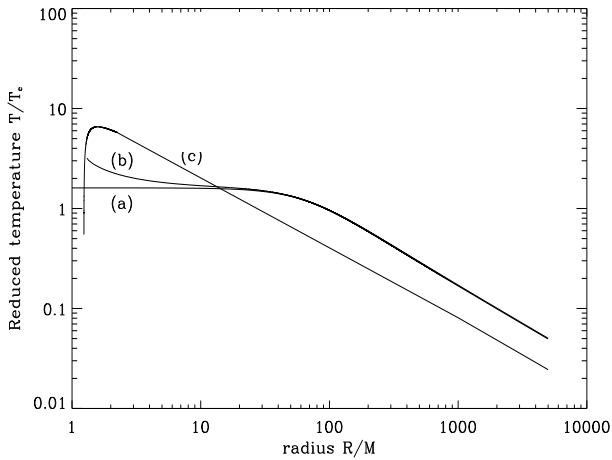}
 \caption[]{\em Effective temperature versus r for $Z_0/M=70$.\\
                            a) Our model in Newtonian metrics\\
                            b) Our model in Kerr metrics\\
                            c) Standard accretion disk 
 }\label{temperature}
\end{figure}

The radiative balance between the hot source and the disk allows to
compute the temperature profile on the disk surface. It is, in
fact, markedly different from ``standard accretion disk model'' as shown
in Figure \ref{temperature}. Indeed, even if at large distances, all
models give the same asymptotic behavior $T\propto R^{-3/4}$, in the
inner part of the disk, it keeps increasing in ``standard model''
whereas, in our model, for $R\leq Z_0$, the 
temperature saturates around a characteristic value $T_c$.
Indeed, the power radiated by the disk is essentially controlled by the
angular distribution of the hot source $\displaystyle\frac{dP}{d\Omega}$
(cf Eq. \ref{eqF}) which is approximatively constant for $R\leq Z_0$
(i.e. $\theta\simeq\pi /4$). The differences between Newtonian and Kerr
metrics comes only from Gravitational and Doppler shifts, which are only
appreciable for $R\leq 5M$. Thus, unless $Z_0$ is itself small enough,
these shifts concern only a small fraction of the emitting area at
$T=T_c$, and modified hardly the UV to X-ray spectrum.

\subsection{The overall spectra}
\begin{figure*}[htb]
        \psfig{width=17cm,height=12cm,file=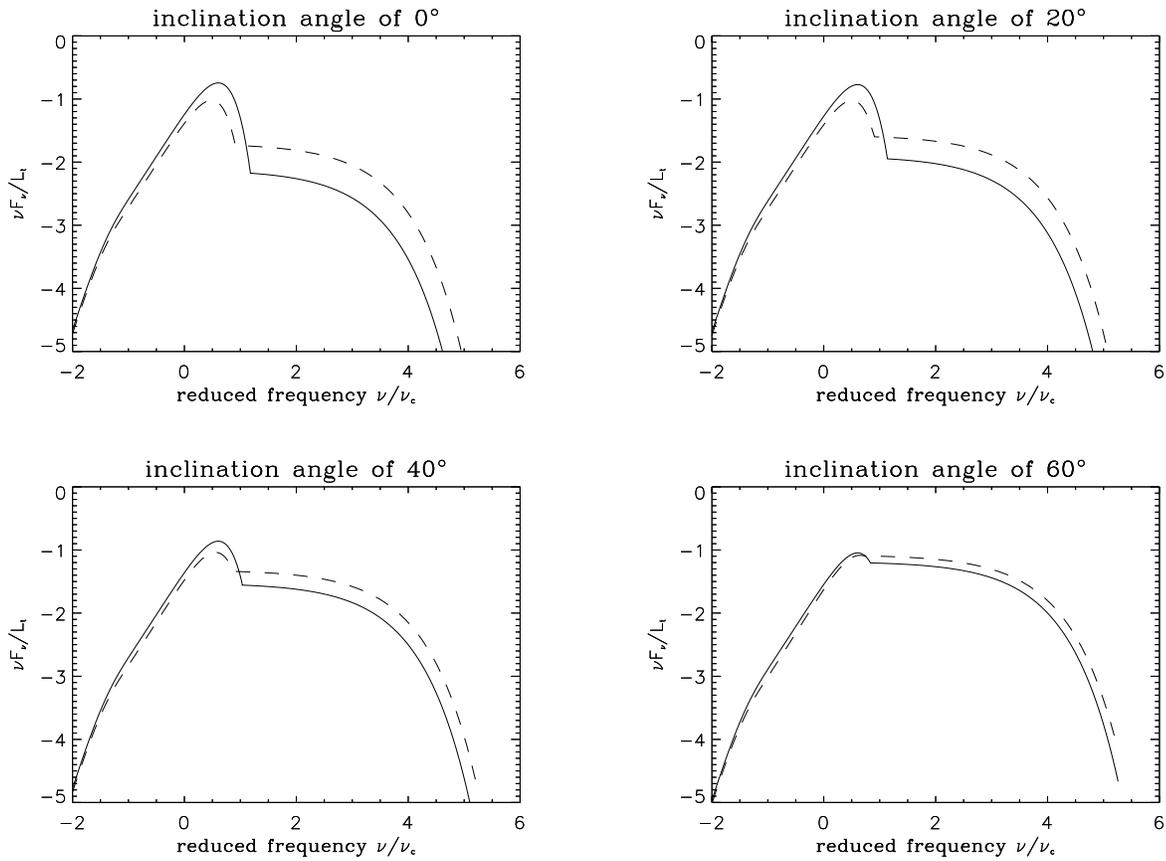}
        \caption[]{\em Differential power spectrum for different inclination
          angle, in the Newtonian (solid lines) and the Kerr maximal
          (dashed lines) cases for $Z_0/M=10$. We use reduced
          coordinates.}\label{specang}  
\end{figure*}
The overall UV to X-ray spectra can be deduced from this model. The bulk of
the energy coming from the disk is emitted on the blue and the
ultraviolet, giving the well-known ``blue-bump'' observed in most quasars
and many AGNs. On the other hand, the high energy spectrum is a power law
with a exponentially cut-off. It depends directly on the relativistic
particle distribution adopted.

\subsubsection{Influence of the inclination angle}
One can see on Figure \ref{specang} Newtonian and Kerr maximal spectra
for different inclination angles for $Z_0/M=10$. For all
inclination angles, the Kerr spectra are always weaker in UV and
 brighter in X-ray than the Newtonian ones. However, the difference tends
to be less visible for the highest inclination angles. This can be
explained easily by the fact that, for high inclination angle, the part
of the disk moving toward the 
observer emits blue-shifted radiation, compensated by the red-shifted
radiation from the other parts. These effects are much less pronounced
for high $Z_0/M$ values because the emission area is much larger, and
thus is less affected by relativistic corrections.  

\subsubsection{Influence of the hot source height}
\begin{figure}
        \psfig{width=8.cm,height=7.cm,file=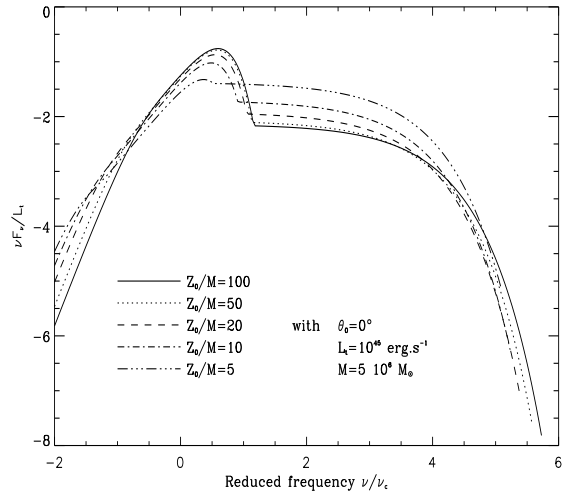}
        \caption[]{\em Differential power spectrum for different values
          of $Z_0$ for the Kerr maximal case. We use reduced
          coordinates.}\label{speczovar} 
\end{figure}

Figure \ref{speczovar} shows the overall spectrum, for
different values of $Z_0/M$. The relativistic effects become important
for values of 
$Z_0/M$ smaller than about $50$. They produce a variation of intensity
lowering the blue-bump and increasing the hard X-ray emission. The change
in the UV range is due to the transverse Doppler effect between the rotating
disk and the observer, producing a net red-shift. In the X-ray range,
the variation is due to the high energy dependence on $Z_0/M$ (cf Figure
\ref{anisotropy}). The observed X/UV 
ratio can then be strongly altered by these effects. Quantitatively, the 
luminosity ratio between the maximum of the blue-bump and the X-ray
plateau goes from $\simeq 30$ in the Newtonian case, to $\simeq 1.5$ for 
$Z_0/M=5$.

\subsection{Scaling laws}
With some further assumptions, the model predicts scaling laws quite different
from the standard accretion models. If one assumes a constant high energy
cut-off (possibly fixed by the pair production threshold) and a
constant solid angle subtended by the hot source, then the 
  following mass scaling laws apply:
\begin{eqnarray}
        T_{c} & = & constant\nonumber  \\
        L_{c} & \propto & M^{2} \nonumber
\end{eqnarray}
That is, the disks have all the same central temperature,
independent of the mass, and a luminosity varying like the mass squared.
 
\section{CONCLUSIONS}
We have developed a new model for Seyfert galaxies emission, which
offers an alternative picture to the standard accretion  
disk emission law. It reproduces individual spectra in agreement with
observations, and predicts new scaling laws that fit better the observed
statistical  
properties. A more complete work has to be done to explain the exact
mechanism of emission of the hot source, supposed to be realized by a
strong shock 
terminating an aborted jet.


\begin{thebibliography}{}

\bibitem[\protect\astroncite{Clavel et~al.}{1992}]{Cla92}
Clavel, J., Nandra, K., Makino, F., Pounds, K. A., Reichert, G. A., Urry,
C. M., Wamsteker, W., Peracaula-Bosch, M., Stewart, 1992, ApJ, 393, 113

\bibitem[\protect\astroncite{Ferreira \& Pelletier}{1995}]{Ferr}
Ferreira, J., Pelletier, G., 1995, A\&A, 295, 807

\bibitem[\protect\astroncite{Ghisellini et~al.}{1991}]{Ghi91}
Ghisellini, G., George, I. M., Fabian, A. C., Done, 1991, MNRAS, 248, 14

\bibitem[\protect\astroncite{Henri \& Petrucci}{submitted}]{Hen*}
Henri, G., Petrucci, P.O., 1997, submitted

\bibitem[\protect\astroncite{Petrucci \& Henri}{submitted}]{Pet*}
Petrucci, P.O., Henri, H., 1997, submitted

\bibitem[\protect\astroncite{Pounds et~al.}{1990}]{Pou90}
Pounds, K. A., Nandra, K., Stewart, G. C., George, I. M., Fabian, 1990, Nature,
344, 132

\end{thebibliography}
\end{document}